\documentclass[pre,reprint,amsmath,amssymb,showpacs]{revtex4-1}
\usepackage{dcolumn}
\usepackage{bm}
\usepackage[dvips]{color}
\usepackage{subfigure}
\usepackage{graphicx} 

\newcommand{\rvec}{\textit{\textbf{r}} }

\begin{document}

\title{Anomalous diffusion of proteins in sheared lipid membranes}

\author{Atefeh Khoshnood$^{1,2}$}

\author{Mir Abbas Jalali$^{1}$}
\email{mjalali@sharif.edu}
\affiliation{$^1$Computational Mechanics Laboratory, Department of Mechanical Engineering,
Sharif University of Technology, Azadi Avenue, Tehran, Iran.\\
$^2$Center of Excellence in Design, Robotics and Automation, Department of Mechanical Engineering,
Sharif University of Technology, Azadi Avenue, Tehran, Iran.}


\begin{abstract}
We use coarse grained molecular dynamics simulations to investigate diffusion properties of 
sheared lipid membranes with embedded transmembrane proteins. In membranes without proteins, 
we find normal in-plane diffusion of lipids in all flow conditions. Protein embedded membranes 
behave quite differently: by imposing a simple shear flow and sliding the monolayers of the 
membrane over each other, the motion of protein clusters becomes strongly superdiffusive in 
the shear direction. In such a circumstance, subdiffusion regime is predominant perpendicular to 
the flow. We show that superdiffusion is a result of accelerated chaotic motions of protein--lipid 
complexes within the membrane voids, which are generated by hydrophobic mismatch or the 
transport of lipids by proteins. 
\end{abstract}


\maketitle

\section{Introduction}
Lipid bilayers are the essential parts of any living cell. They constitute the main body of 
the cell membrane while being found in different organelles inside the cell. The cell membrane 
hosts collections of proteins and lipid rafts and it is crowded with a variety of biomolecules. 
In such non-homogenous and diverse environments, the diffusion of protein molecules in lipid 
bilayers plays a vital role in different biological processes like cell signaling. The diffusion of 
lipids and proteins is not a distinct phenomenon and depends on the environment and 
neighboring molecules \cite{dix08}, and even changes from cell to cell \cite{wie09}.
Transmembrane proteins diffuse as dynamic complexes with lipids \cite{nie10,pra07}, 
and their interactions with lipid molecules mediates traffic in cell membranes. Experiments 
show that the hydrophobic mismatch between proteins and lipids controls the diffusion 
coefficient of molecules inside a bilayer \cite{ram10,gam10}. 

Anomalous sub- and super-diffusion processes are more efficient scenarios for finding 
a nearby target than normal diffusion \cite{gui08,bar02}, and they enhance the formation 
of protein complexes and signal propagation. According to experiments, the mean square 
displacement (MSD) of membrane channel proteins of human kidney cell exhibits 
subdiffusion \cite{wei11}. The addition of cholesterols to lipid membranes \cite{jeo12}, 
and the augmented area coverage of membrane proteins \cite{ram09} also lead to 
subdiffusion of lipids and proteins. Superdiffusivity has been observed in several 
physical systems and is often associated with L\'evy flights. Prominent examples 
are the chaotic motion of particles in a rotating laminar flow \cite{sol93} with long-range
flights and horizontally vibrated grains which exhibit L\'evy flight with small jumps 
compared to their diameter \cite{lec10}. Nevertheless, an active component such as 
molecular motor can also be the source of superdiffusivity. A recent study by \citet{koh11} 
shows that in a gel composed of actin filaments, fascin molecules and myosin-II filaments, 
the diffusion of small actin and fascin clusters are superdiffusive because of the work 
done by molecular motors.

In many conditions membranes are under shear. When a red blood cell (RBC) 
migrates through vessels smaller in diameter than itself, the RBC membrane is under 
shear. The blood flow exerts tangential shear stresses on vascular endothelia, and 
initiates cellular processes like activating G protein-coupled receptors. These receptors 
are able to sense the fluid shear stress as an increase in the lateral membrane tension, 
and subsequently go through conformational changes \cite{cha06}. 
The temporal and spatial changes in the membrane fluidity, in response to shear 
flow, have been observed experimentally \cite{but01,hai00}. 

In this study we are 
interested in the diffusivity of lipid and protein molecules in flat membranes under 
shear flow, and attempt to answer three fundamental questions using molecular 
dynamics (MD) simulations: (i) Do lipid molecules have different diffusion coefficients 
parallel and perpendicular to flow direction? (ii) How does a simple shear flow 
influence the random motions of transmembrane proteins? (iii) Is there any correlation 
between the population of proteins and their diffusion in the membrane? \\
\section{Model and methods}
\begin{figure}[t]
\centerline{\hbox{\includegraphics[trim = 10mm 0mm 0mm 0mm,width=0.23\textwidth]{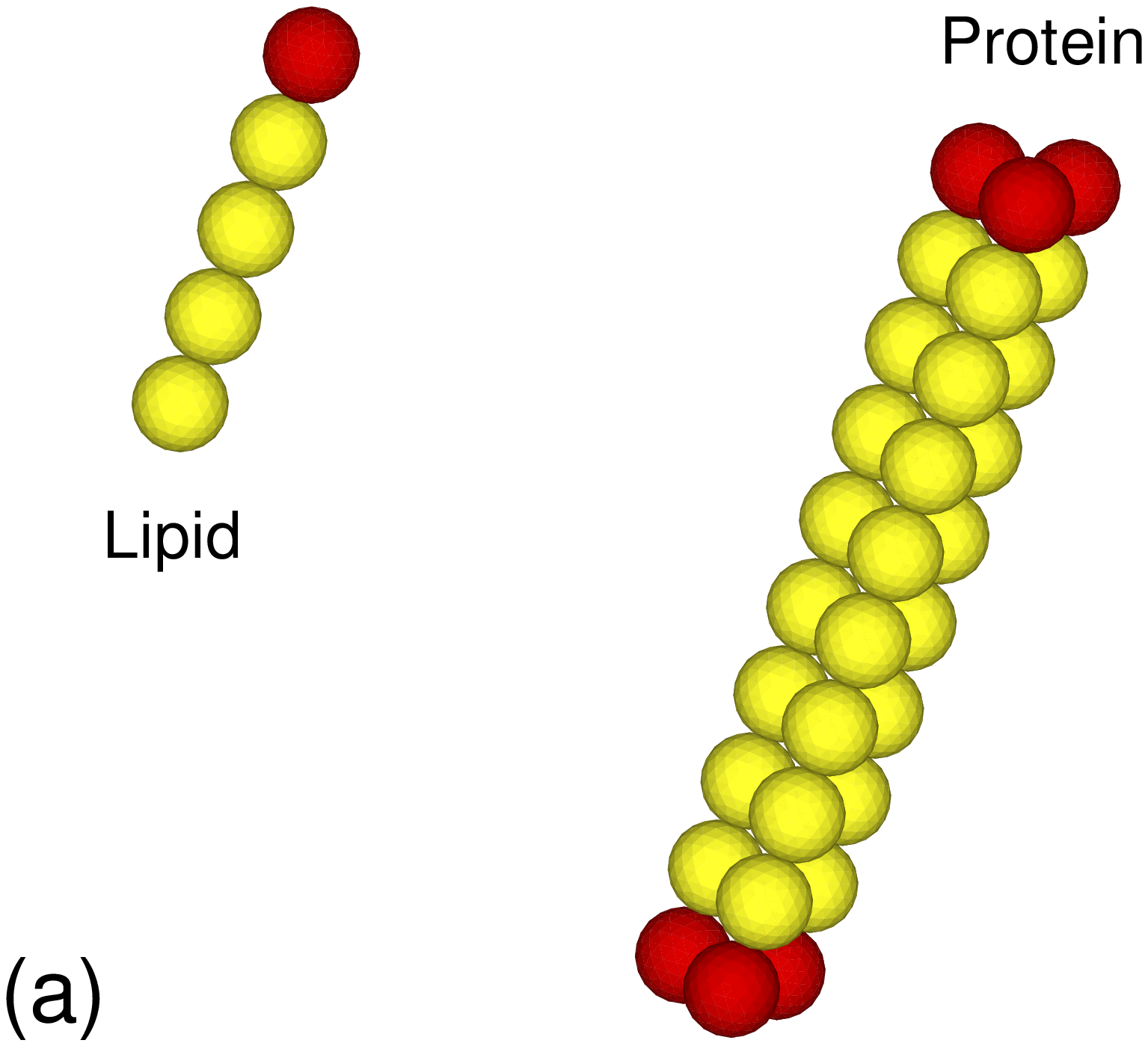}}
                    \hbox{\includegraphics[trim = 10mm 0mm 0mm 0mm,width=0.23\textwidth]{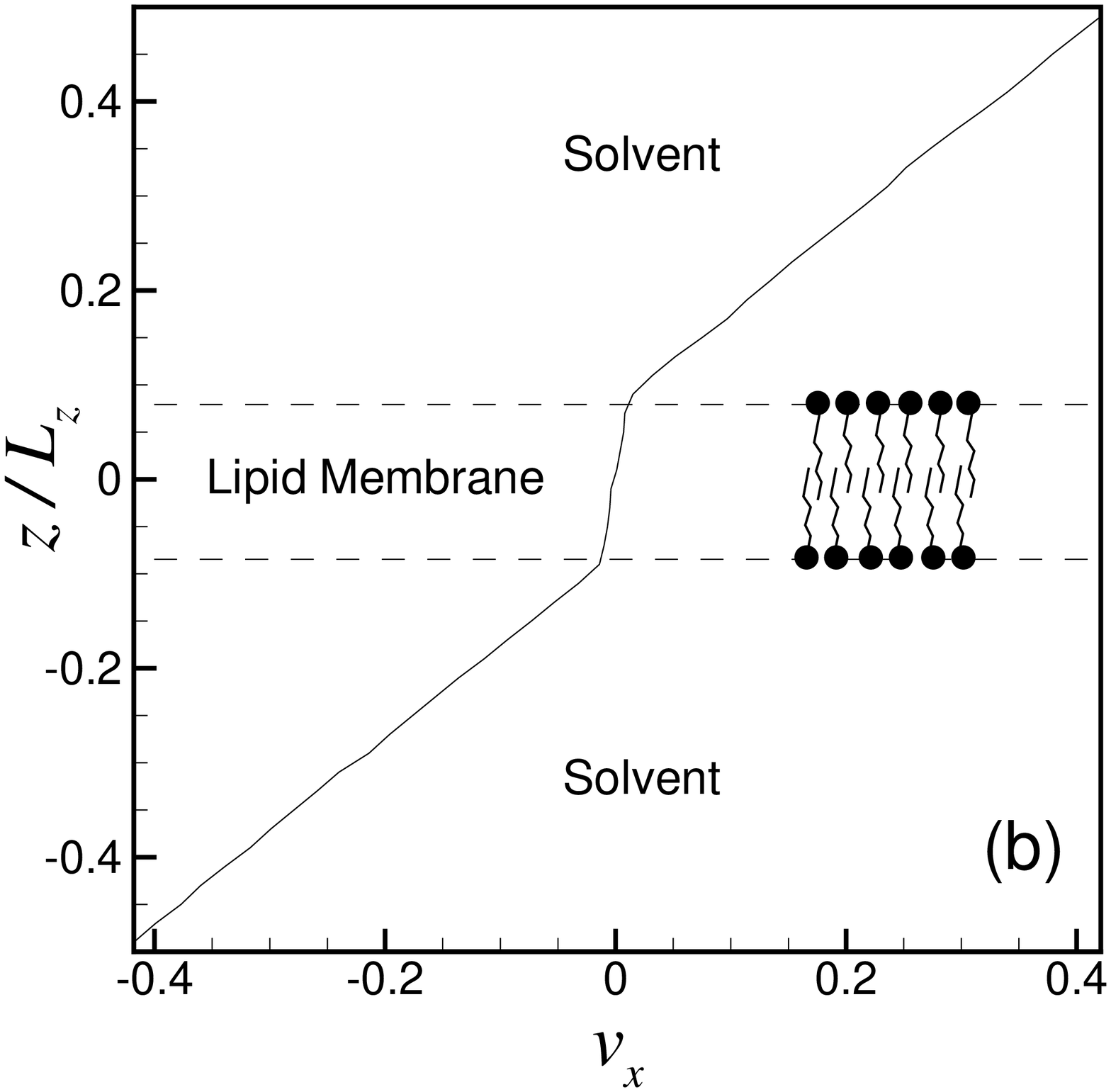}} }
\caption{(Color online)(a) The models of lipid and protein molecules. Red and yellow spheres are hydrophilic
and hydrophobic particles, respectively. (b) Velocity profile of a sample solvent--membrane 
system under simple shear flow with $\dot\gamma=0.03\tau^{-1}$. The dashed lines mark 
the average boundaries of the solvent columns and the membrane.}
\label{fig:vprofile}
\end{figure}

We simulate lipid membranes utilizing a flexible lipid model \cite{GL} and 
triple-strand rigid proteins \cite{khosh10} [Fig. \ref{fig:vprofile}(a)]. Although
different coarse grained models have been developed over years \cite{mar07}, 
the model adopted here has the ability to mimic the physical properties of lipid 
membranes. The model is not a true coarse grained model but can qualitatively describe 
phenomena related to lipid membranes and associated transmembrane proteins as we will 
compare some of our results with true coarse grained and atomistic models. Our 
goal is to explore the effect of shear flow on the motion of lipid and protein molecules 
over nanosecond time scales. 
We perform MD simulations of an $NVT$ ensemble, where the number of particles $N$, the 
volume $V$ and the temperature $T$ are held constant. Lees-Edwards boundary condition is 
employed to generate simple shear flow with the shear rate $\dot\gamma$ \cite{AT}. Other 
boundary conditions are periodic. The temperature is set to $324 \, {\rm K}$ so that the 
system is safely above the gel to liquid phase transition temperature of different 
phosphatidylcholine lipid bilayers. A detailed description of the model can be found 
in \citet{khosh10}.

We express the position and velocity vectors of particles in the Cartesian $(x,y,z)$ 
coordinate system whose origin is located at the center of our cubic simulation box. 
The $x$ and $y$ axes lie in the membrane plane and the $z$ axis is perpendicular 
to that. MD scales of length, time, mass and energy are $\sigma=1/3~{\rm nm}$, 
$\tau=1.4~{\rm ps}$, $N_{\rm avo}m=36~{\rm g/mol}$ and $N_{\rm avo}\epsilon=2~{\rm kJ/mol}$, 
respectively. $N_{\rm avo}$ is the Avogadro's number. In all simulations, the dimensions 
of the box along the coordinates axes, $L_x$, $L_y$ and $L_z$, are set to 
$L_x=L_y=L_z=28.71\sigma$. The total number of particles equals $N=15625$, 
which gives a fixed number density, $\rho= 2/(3\sigma^3)$, and a constant average 
fluid pressure of $(1.7 \pm 0.1) \epsilon\sigma^{-3}$ for all simulations. Here, we note that 
isotropic pressure control is not appropriate in the simulation of lipid membranes since their 
volume is constant in the laboratory and biological conditions. However, a fixed number density 
will give a constant average pressure and physical properties of lipid membranes with different 
number of lipids and proteins can be compared. Physical quantities are measured 
using a run-time of $\approx 5000\tau$. An important mechanical property of every membrane is the 
surface tension $\zeta$ that mainly affects the diffusion of lipid molecules. We compute the 
tension of our model membranes from
\begin{equation}
\zeta=\left[ P_{zz}-\left( P_{xx}+P_{yy}\right)/2 \right]L_z
\end{equation}
where $P_{\alpha\alpha}$ ($\alpha\equiv x,y,z$) are the components of the pressure tensor \cite{AT}. 

In this study we apply MSD to determine the diffusion properties of randomly moving particles.
The diffusion coefficient is thus calculated using the Einstein expression
\begin{equation}
D_{\alpha\alpha}=\lim_{t\to\infty}\frac{1}{2Nt}\langle\sum^N_{i=1}\left[q_{i\alpha}(t)-q_{i\alpha}(0)\right]^2\rangle,
\label{eq:class-diff}
\end{equation}
where $\alpha\in\left\lbrace x,y,z\right\rbrace$ and $q_{i\alpha}$ is the displacement 
due to the random motion of the $i$th particle in the $\alpha$-direction. The summation 
in Eq. (\ref{eq:class-diff}) is taken over the particles of the same type. From here on, we 
will drop the summation sign for brevity. The operator $\langle\cdots\rangle$ denotes 
the canonical average. Eq. (\ref{eq:class-diff}) describes the regular Brownian motion 
when the MSD is linearly proportional to $t$. The diffusion process is anomalous should 
the MSD deviate from the linear form, and obey the relation
\begin{equation}
\langle\left[q_{\alpha}(t)-q_{\alpha}(0)\right]^2\rangle=2D^a_{\alpha\alpha}t^a,
\label{eq:class-diff-2}
\end{equation}
where $a$ is the diffusion exponent and $D^a_{\alpha\alpha}$ is the fractional 
diffusion coefficient. The regimes with $0<a<1$ and $a>1$ are subdiffusive
and superdiffusive, respectively. To obtain smooth MSD curves, we evolve systems 
of 80 different initial conditions and report their ensemble-averaged diffusion coefficients
and MSDs. The diffusion of lipids is investigated by tracing the motion of their head groups. 
Proteins are traced using their center of masses. 

In equilibrium models without external shearing, there is no streaming in the 
solvent--membrane system. Therefore, the flux of particles is associated with their random 
motions, and any displacement is due to thermal fluctuations. In sheared membranes, 
however, there is a combination of streaming and diffusive fluxes. We thus need to 
distinguish and eliminate the streaming flux when calculating the MSD. Let us define 
the actual velocity components of the $j$th particle as 
$v_{j\alpha}=\langle v_{\alpha} \rangle +\tilde v_{j\alpha}$ where $\langle v_{\alpha} \rangle$ 
is the average streaming velocity, and $\tilde v_{j\alpha}$ is the peculiar velocity 
whose time integral gives the displacements in Eqs. (\ref{eq:class-diff}) and (\ref{eq:class-diff-2}). 
In equilibrium models, the average velocities $\langle v_{\alpha}\rangle$ vanish and we 
obtain $q_{j\alpha}=\int v_{j\alpha} dt=\int \tilde v_{j\alpha} dt$. With external shearing, the flow 
is always imposed in the $x$-direction. Therefore, $v_{jy}=\tilde v_{jy}$ and $v_{jz}=\tilde v_{jz}$ 
are directly integrated to find the corresponding displacements. When the simulation box is 
uniformly filled with one type of particles (let us says solvent particles), one readily finds 
$\tilde v_{jx}=v_{jx}-z \dot\gamma$. In the presence of a lipid bilayer, the vertical velocity 
profile in the $z$-direction is no longer linear [see Fig. \ref{fig:vprofile}(b)]. Therefore, to obtain 
the MSD of lipids, we define $\langle v_{x} \rangle$ as the average velocity of the layer 
where the head groups of phospholipds reside, and obtain 
$q_{jx}=\int \left ( v_{jx} - \langle v_{x} \rangle \right ) dt$. 
It is remarked that transmembrane proteins do not experience streaming movements,
$\langle v_{\alpha} \rangle =0$, because the shear forces exerted on their end points 
from the upper and lower solvent columns are equal and in opposite directions.\\
\section{Diffusion of lipids} 
\begin{figure}
\centerline{\hbox{\includegraphics[width=0.4\textwidth]{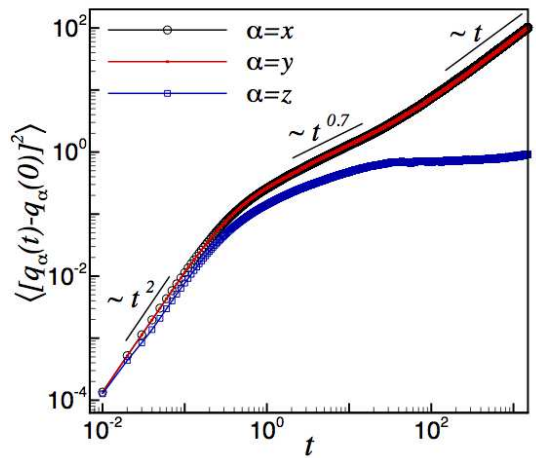} } }
\caption{(Color online) MSD of lipid molecules for the membrane with 600 lipids 
and for $\Delta t=0.01\tau$. The membrane is under simple shear flow with $\dot\gamma=0.03\tau^{-1}$, 
and each lipid molecule has been hydrated by almost 20 solvent particles. The coordinate 
axes are in logarithmic scale.}
\label{fig:MSD-l}
\end{figure}

The lateral diffusion of lipids in equilibrium conditions  
is enhanced as the membrane tension increases. This has been observed in 
simulations by atomistic \cite{mud11} and coarse grained models \cite{neder10} and is because 
by stretching the membrane, the area per lipid increases and more space is provided for 
the free motion of lipids. Our simulations with this very simple model shows the same pattern. 
By turning on the shear flow, lipid molecules undergo an initial 
ballistic motion that transforms into an interval of subdiffusion with $a=0.7$ [Fig. \ref{fig:MSD-l}]. 
The transient anomalous state has been observed in atomistic simulations \cite{fle09} as well.
After the transient anomalous diffusion and over longer time scales a normal diffusion 
with $a=1$ is observed [Fig. \ref{fig:MSD-l}]. It is noted that we have found similar MSD 
profiles for lipid molecules in equilibrium and sheared systems, and in both cases lipids 
ultimately develop normal diffusion. Although \citet{kne11} reported a permanent subdiffusive 
behavior by the atomistic simulation of lipid membranes in equilibrium, experiments support a 
final regular diffusion regime, as we do, even in the presence of obstacles \cite{ska11}. 
We conclude that the diffusion regime of lipid molecules is invariant with and without 
external shearing.

The diffusion coefficients obtained from the normal diffusion region of MSD plots, 
are larger for smaller shear rates. For example, for a membrane 
of $N_l=600$ lipid molecules, we find $\zeta=(1.4846 \pm 0.2624)\epsilon/\sigma^2$, 
$D_{xx}=0.0338\sigma^2/\tau$ and $D_{yy}=0.0334\sigma^2/\tau$. For the same 
system under a shear flow of $\dot\gamma=0.03\tau^{-1}$, the membrane tension 
drops to $\zeta=(0.8163 \pm 0.2727)\epsilon/\sigma^2$ and diffusion coefficients 
reduce to $D_{xx}=0.0318\sigma^2/\tau$ and $D_{yy}=0.0332\sigma^2/\tau$. 
The reason is that the membrane thickness increases for higher shear rates and the 
tension decreases without any change in the area per lipid \cite{khosh10}. Consequently, 
the fluidity of the membrane decreases and slows down the diffusion process. 
After applying the shear force, we find that $D_{xx}$ drops for about $6\%$ while 
$D_{yy}$ remains almost constant with only $0.4\%$ change which is not statistically 
significant since it is less than the mean standard error for diffusion coefficients 
which is less than $1\%$. The difference between $D_{xx}$ and $D_{yy}$ is indistinguishable in 
Fig. \ref{fig:MSD-l}. We speculate that the alignment of lipid chains with the flow 
breaks the isotropy and yields $D_{xx} \not =D_{yy}$. Atomistic simulation of lipid membrane 
\cite{mud11} has shown that increasing the tension of membrane, induced by altering the area 
per lipid, result in larger lateral diffusion coefficients. This change is not linear with 
tension and they have reported $4-28\%$ increase in lateral diffusion coefficient. Coarse grained 
simulations \cite{neder10} showed that for larger tensions the increase in diffusion 
coefficient slows down and it depends on the range of tension. In our simulation, shear flow 
induces $45\%$ change in tension and consequently result in a different diffusion coefficient 
in the flow direction. In the $z$-direction, perpendicular to the membrane plane, our MSD 
plots always show a confined motion as is expected.\\
\section{Diffusion of proteins} 
We add rod-like proteins to the membrane, and simulate models with different protein 
concentrations that vary significantly from cell to cell. Since proteins increase the membrane 
tension as they perturb the distribution of lipids \cite{khosh10}, we increase the number of 
lipids (proportional to proteins) to keep the membrane almost tensionless. In equilibrium and for 
a membrane with a single embedded protein with 
$\zeta=(0.1153\pm0.1742)\epsilon/\sigma^2$, we can measure the diffusion coefficients 
$D_{\alpha\alpha}$ since the MSD of protein shows an ultimate regular diffusion. 
We find $D_{xx}=0.0253\sigma^2/\tau$ and $D_{yy}=0.0254\sigma^2/\tau$, which are 
equivalent to $D_{xx}\approx D_{yy}\approx 2\times10^{-9}~{\rm m^2/s}$ with less than $1\%$ error. 
These values are larger than experimental values, by two orders of magnitude. The obvious reason is 
the effect of coarse graining that has reduced the interdigitation and friction between molecules,
and allows for faster movements of particles. The reduced friction affects both membrane component 
and solvent motion and consequently decreases both solvent and membrane viscosity. Viscosity of a 
fluid is a determinant of mobility or diffusion in that medium. Another minor 
source of discrepancy is the smaller size of our model proteins compared to real integral proteins. 
\begin{figure*}
\centerline{\hbox{\includegraphics[width=0.34\textwidth]{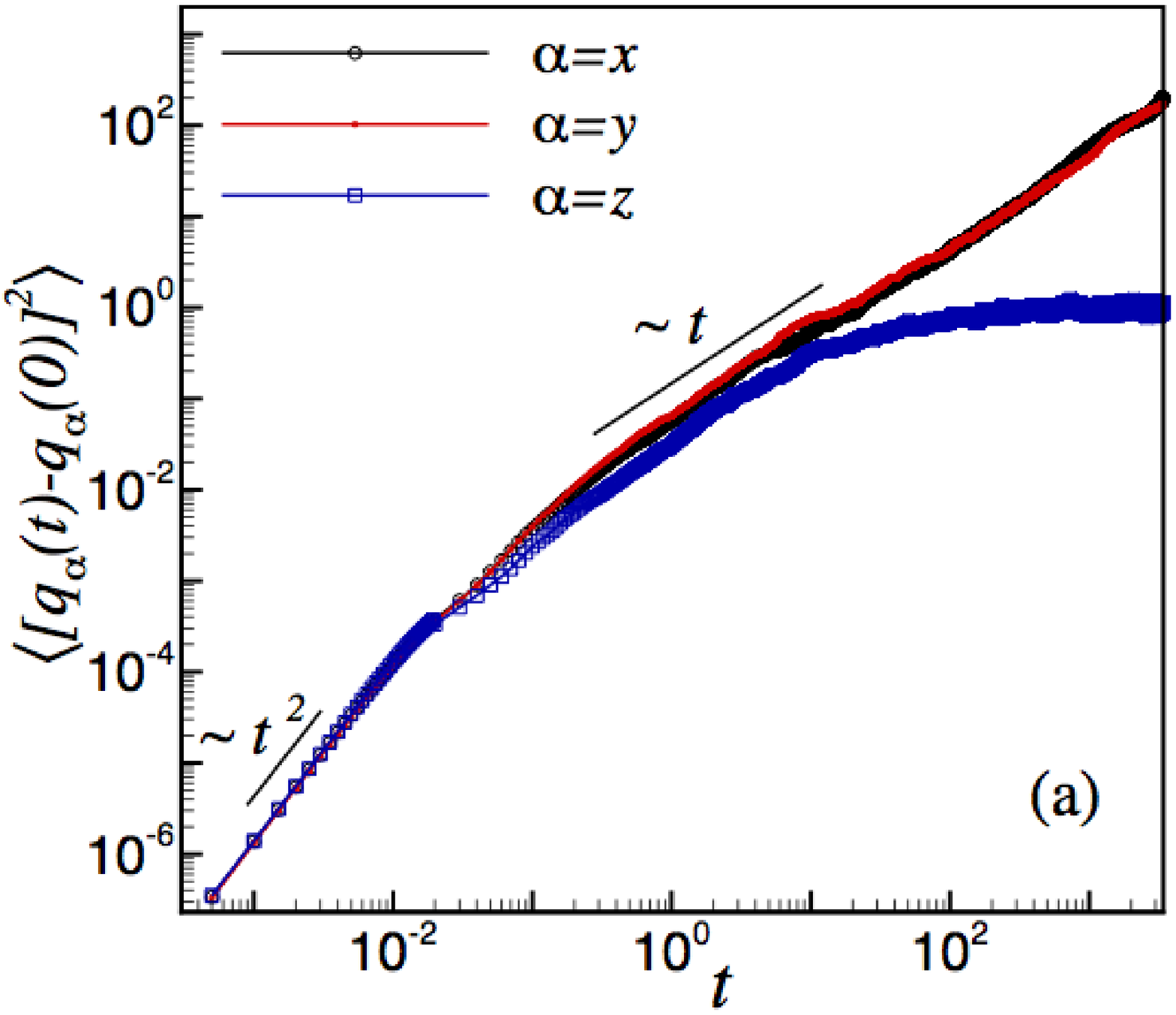} }  \hspace{-0.2in}
            \hbox{\includegraphics[width=0.34\textwidth]{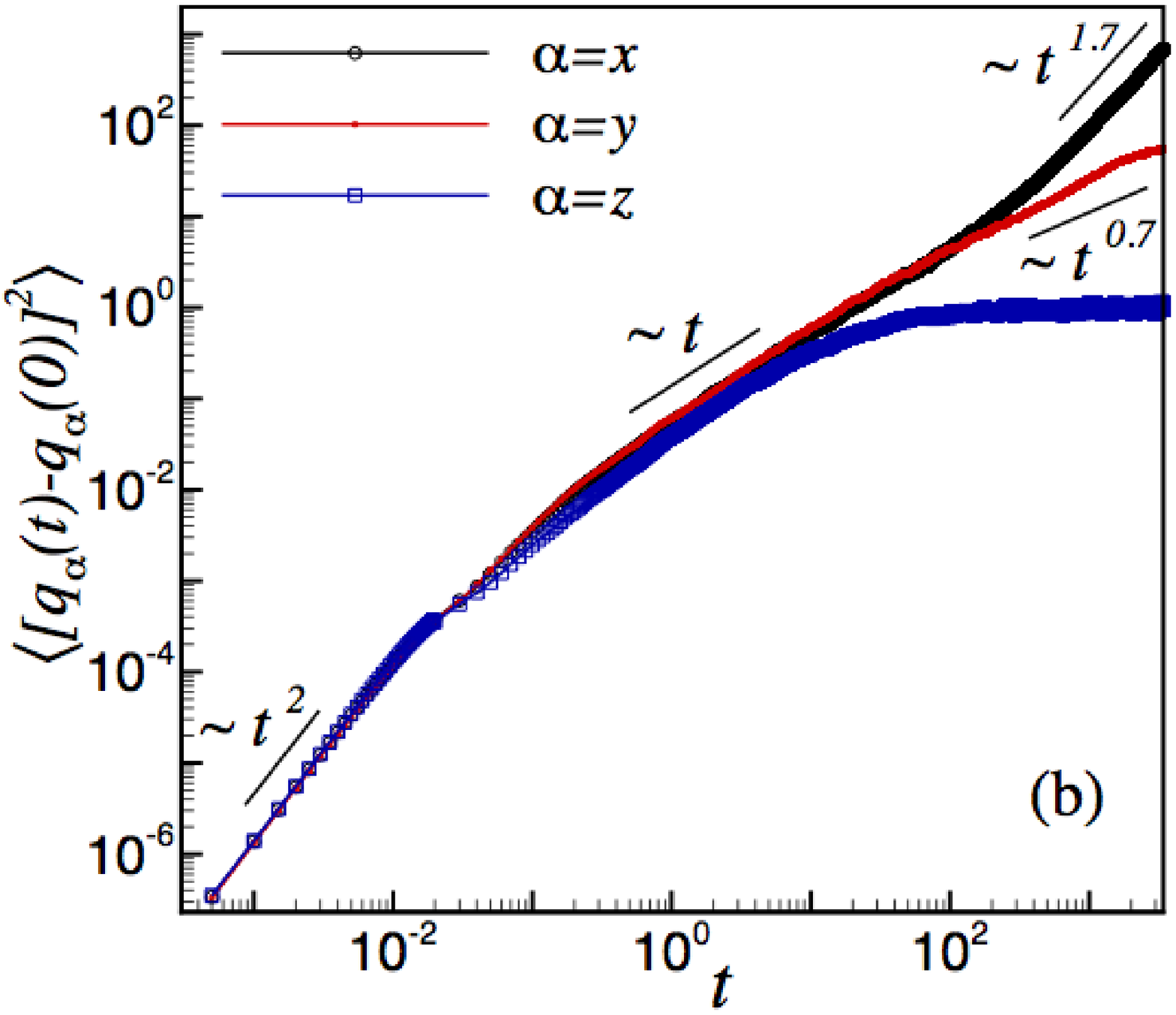} }  \hspace{-0.2in} 
            \hbox{\includegraphics[width=0.34\textwidth]{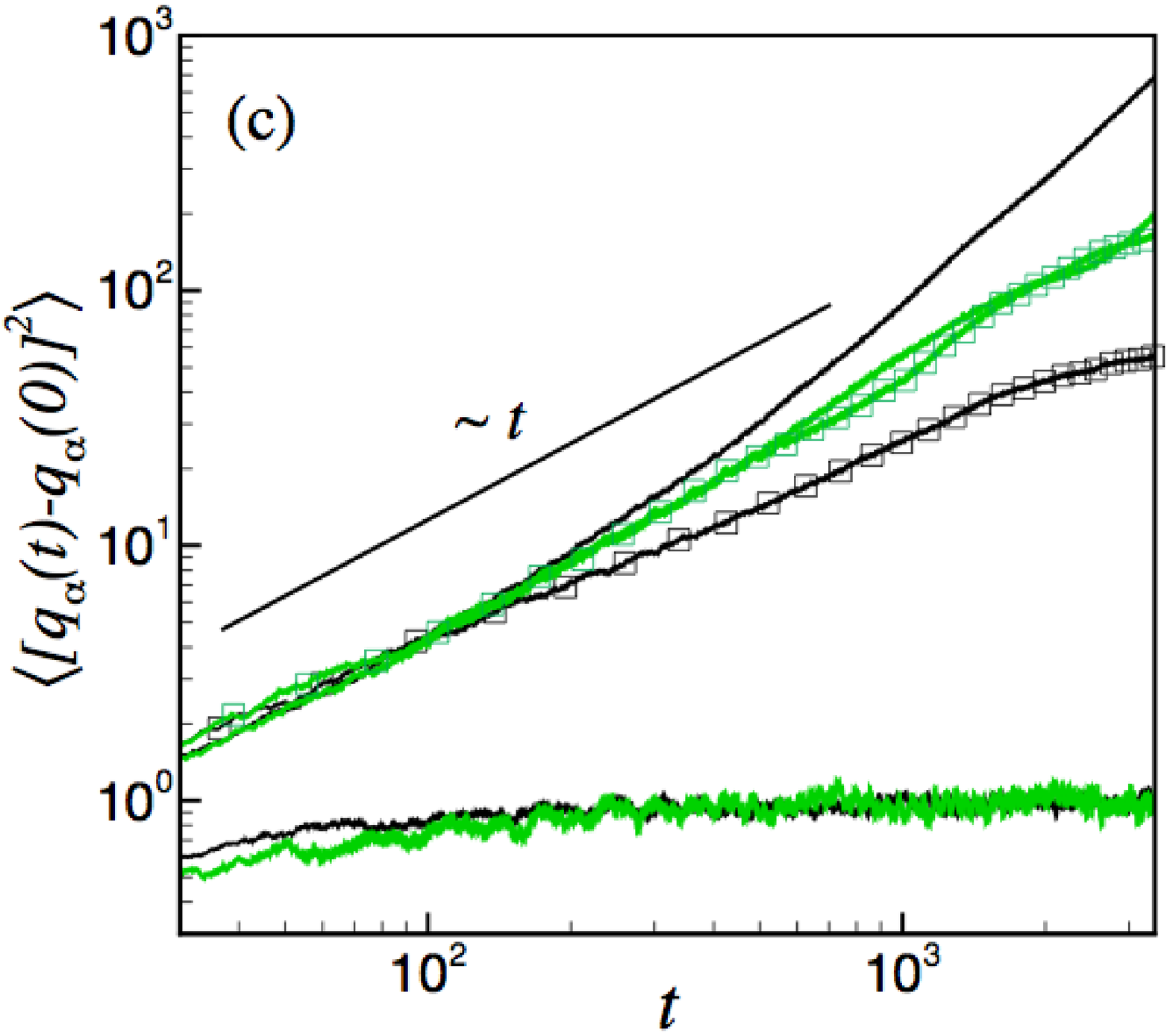} } }
\caption{(Color online) (a) MSD of protein 
molecules in a protein-embedded model with 2 and (b) 4 proteins, which correspond to 640 
and 660 lipids, respectively. We have used $\Delta t=0.0005\tau$ up to $t=0.02\tau$, and 
$\Delta t=0.01\tau$ for $t>0.02\tau$. In both panels (a) and (b) 
the membrane is under simple shear flow with $\dot\gamma=0.03\tau^{-1}$, and each lipid 
molecule has been hydrated by almost 20 solvent particles. The coordinate axes are in 
logarithmic scale. (c) Zoomed upper parts of the MSD profiles with the green (light gray) 
and black lines corresponding to the systems of panels (a) and (b), respectively. The 
profiles with square symbols correspond to the $y$-direction.}
\label{fig:MSD-p}
\end{figure*}

We note that the average diffusive behavior of lipids is unaffected by the presence of proteins. 
We computed the average MSD profile of lipids for the above system and found the same pattern of 
initial ballistic regime, transient subdiffusive region and final normal diffusion. 
The diffusion coefficients for our model proteins is smaller than model lipid molecules by a factor of $0.75$. 
This is expected because of the larger size and mass of proteins compared to lipids. 

An initial ballistic motion the same as what has been observed for lipids is recovered for proteins 
by using time step as small as $\Delta t=0.0005\tau$ for $t<0.01\tau$. This regime is shared 
by all the systems regardless of the shear rate and the number of proteins.
By putting the system under simple shear flow, proteins undergo Brownian motion when only two 
proteins are used [Fig. \ref{fig:MSD-p}(a)]. One could anticipate this result, for single proteins cannot 
remarkably perturb the distribution of lipids, and change the diffusion properties of the membrane. 

With 4 proteins, however, we observe that they form two double-protein clusters (due to the depletion force), 
and exhibit a strong superdiffusive motion parallel to the flow. Fig. \ref{fig:MSD-p}(b) shows 
how after $t\sim 100\tau$ the normal diffusion regime transforms to strong superdiffusion 
with $a=1.7$ in the $x$-direction. Interestingly, this exponent is the same as the superdiffusion 
exponent found by \citet{koh11} for active diffusion of protein clusters by molecular motors. 
Because of crowding effect and increase in the concentration of proteins \cite{dix08}, our results show 
a subdiffusive behavior along the $y$ axis  with $a=0.7$. \citet{wei11} observed $a=0.8\pm0.1$ 
in experiments with channel proteins of human kidney cell, and \citet{jav13} found 
$a=0.75\pm0.15$ by molecular simulations of aggregating NaK channel proteins. For clarity, 
the upper parts of the MSD profiles in Figures \ref{fig:MSD-p}(a) and \ref{fig:MSD-p}(b) are 
plotted together in Fig. \ref{fig:MSD-p}(c).

\begin{figure}
\centerline{\hbox{\includegraphics[width=0.45\textwidth]{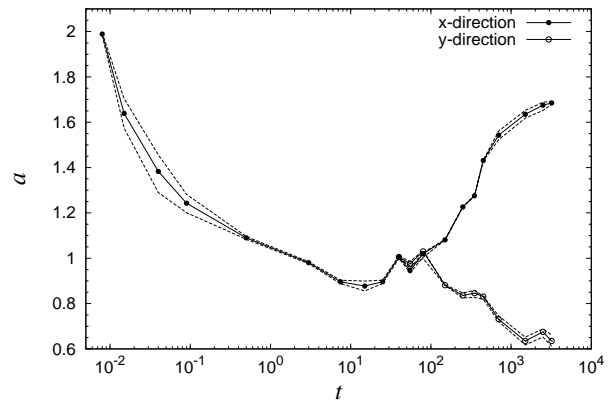} } }
\caption{The variation of the diffusion exponent $a$ for the system of 
Fig. \ref{fig:MSD-p}(b). Solid lines correspond to the mean values and dashed lines show
the error. For the $y$-direction, $a$ has been plotted only for $t>30\tau$ because for 
$t<30\tau$ diffusion exponents of the $x$- and $y$-directions are almost identical. 
The horizontal coordinate axis is in logarithmic scale.}
\label{fig:a}
\end{figure}

To rule out the effect of statistical errors in the development of anomalous behavior, 
we have divided the MSD profiles of the system with 4 proteins to smaller intervals, and 
separately calculated $a$ and its error over each interval using the curve fitting toolbox 
of MATLAB. We have then assigned the mean value to the center of the time interval and 
plotted the calculated diffusion exponent versus time in Fig. \ref{fig:a}. For instance, 
over the initial ballistic zone, we have obtained $a=1.989\pm0.006$ from $t=0.005\tau$ 
to $t=0.01\tau$, and assigned this value to $t=0.0075\tau$. Fig. \ref{fig:a} shows that 
$a$ approaches $1.7$ and $0.7$ in the $x$- and $y$-directions, respectively.

\begin{figure}[t]
\centerline{\hbox{\includegraphics[width=0.4\textwidth]{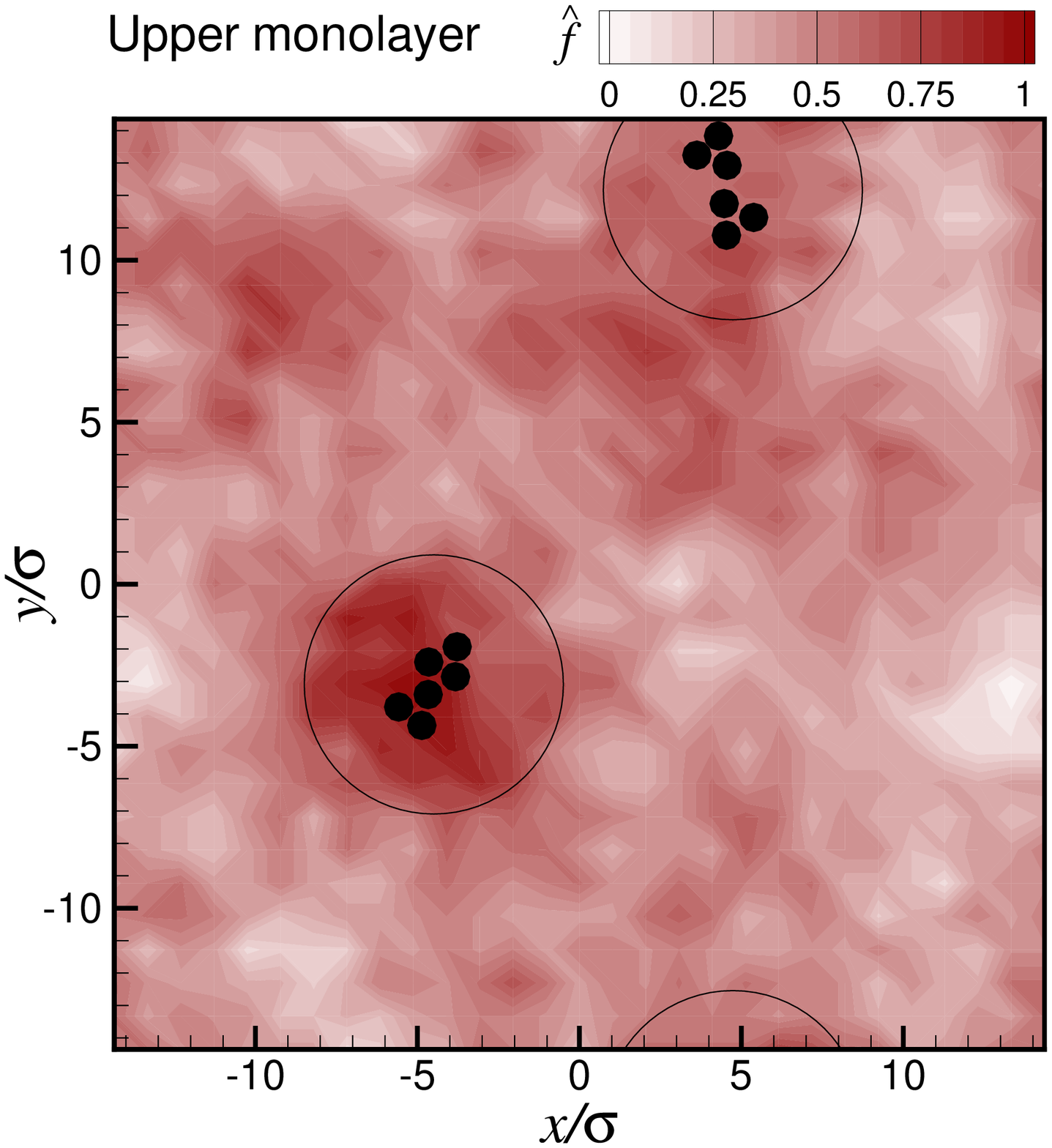} }}
\centerline{\hbox{\includegraphics[width=0.4\textwidth]{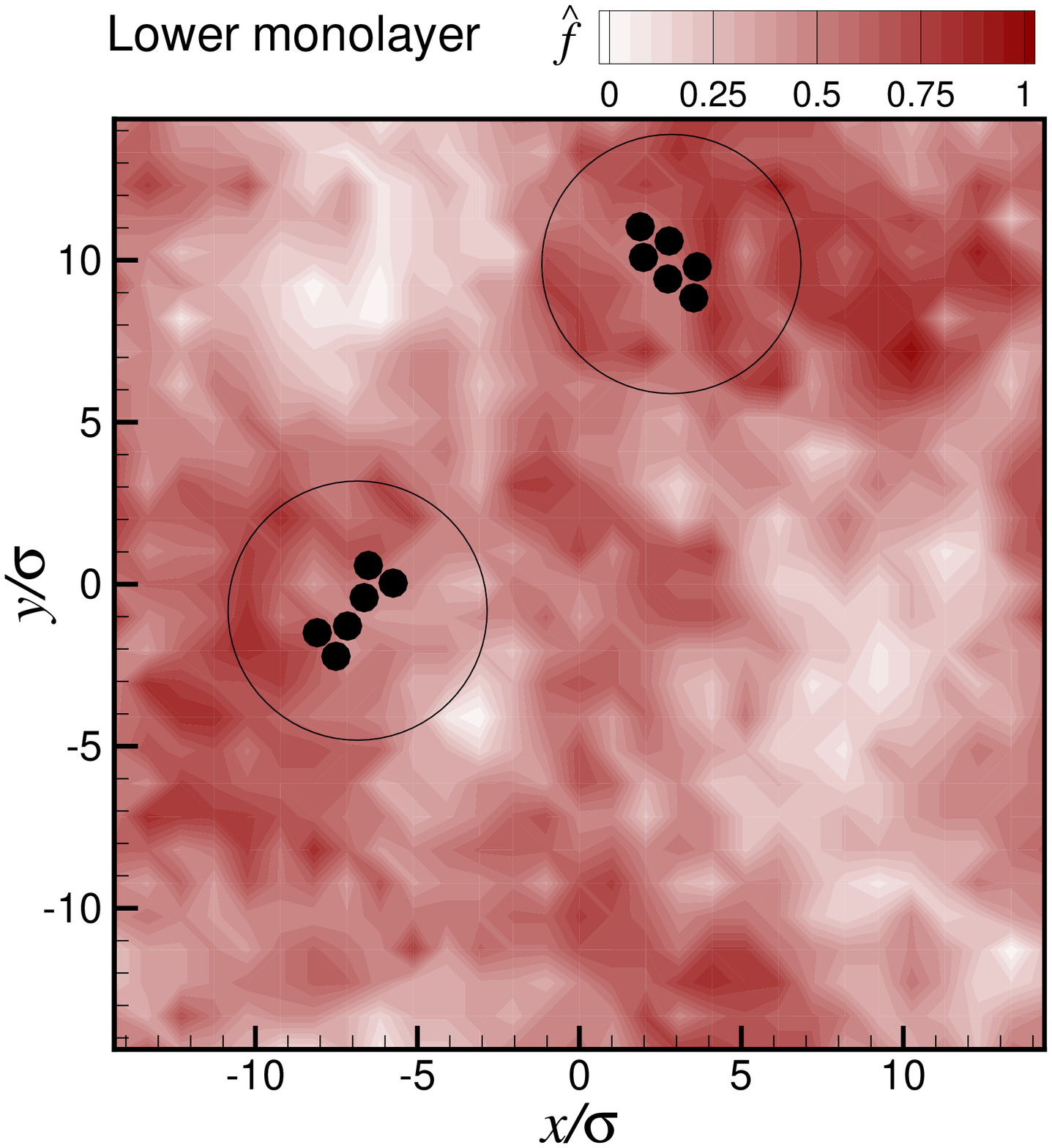} }}
\caption{(Color online) The local concentration $\hat f(\textit{\textbf{r}},\Delta,t)$ of head groups in the 
upper (left panel) and lower (right panel) leaflets of a sheared membrane at a randomly 
selected time for $\Delta=4\sigma$. The head particles of proteins are shown by filled 
circles. Drawn circles have radii of $\Delta$, and their centers lie at the centroid of the 
head particles of proteins. The flow with the shear rate $\dot\gamma=0.03\tau^{-1}$ is 
in the $x^{+}$ and $x^{-}$ directions for upper and lower leaflets, respectively 
[cf. Fig. \ref{fig:vprofile}(b)]. The model has 660 lipids and  two double-protein clusters. 
Since proteins are longer than the bilayer thickness, they are aligned in the shear flow, 
and their upper and lower head particles do not have the same coordinates.}
 \label{fig3}
 \end{figure}

To understand the physical mechanism behind the observed anomaly, we 
conduct the following analysis. Let us define the local concentration 
of the head particles of lipid and protein molecules
at the position $\textit{\textbf{r}}$ and time $t$ as 
$f=\frac {1}{N_{\rm h}} \sum_{i=1}^{N_{\rm h}} \left [ H(\delta_i)-H(\delta_i-\Delta) \right ]$
where $\delta_i(t)=\vert \textit{\textbf{r}}_i(t)-\textit{\textbf{r}} \vert$, and $\textit{\textbf{r}}_i(t)$ 
is the position vector of the $i$th head particle. $N_{\rm h}$ denotes the total number of 
head particles in the monolayer, $H(\xi)$ is the Heaviside step function, and $2\Delta$
is the typical size of the cross section of a protein cluster (or a protein--lipid complex). 
Our numerical experiments show $\Delta=4\sigma$ is the best choice. We examine the 
trajectories of protein molecules and the spatial variation of the normalized distribution 
$\hat f(\textit{\textbf{r}},\Delta,t)=(f-f_{\rm min})/(f_{\rm max}-f_{\rm min})$ to explain 
the physics of observed superdiffusion. Here $f_{\rm min}$ and $f_{\rm max}$ are the 
minimum and maximum values of $f$ at a given time $t$. Fig. \ref{fig3} demonstrates 
contour plots of $\hat f$ for the upper and lower monolayers at a randomly chosen time.

\begin{figure*}
\centerline{\hbox{\includegraphics[width=0.4\textwidth]{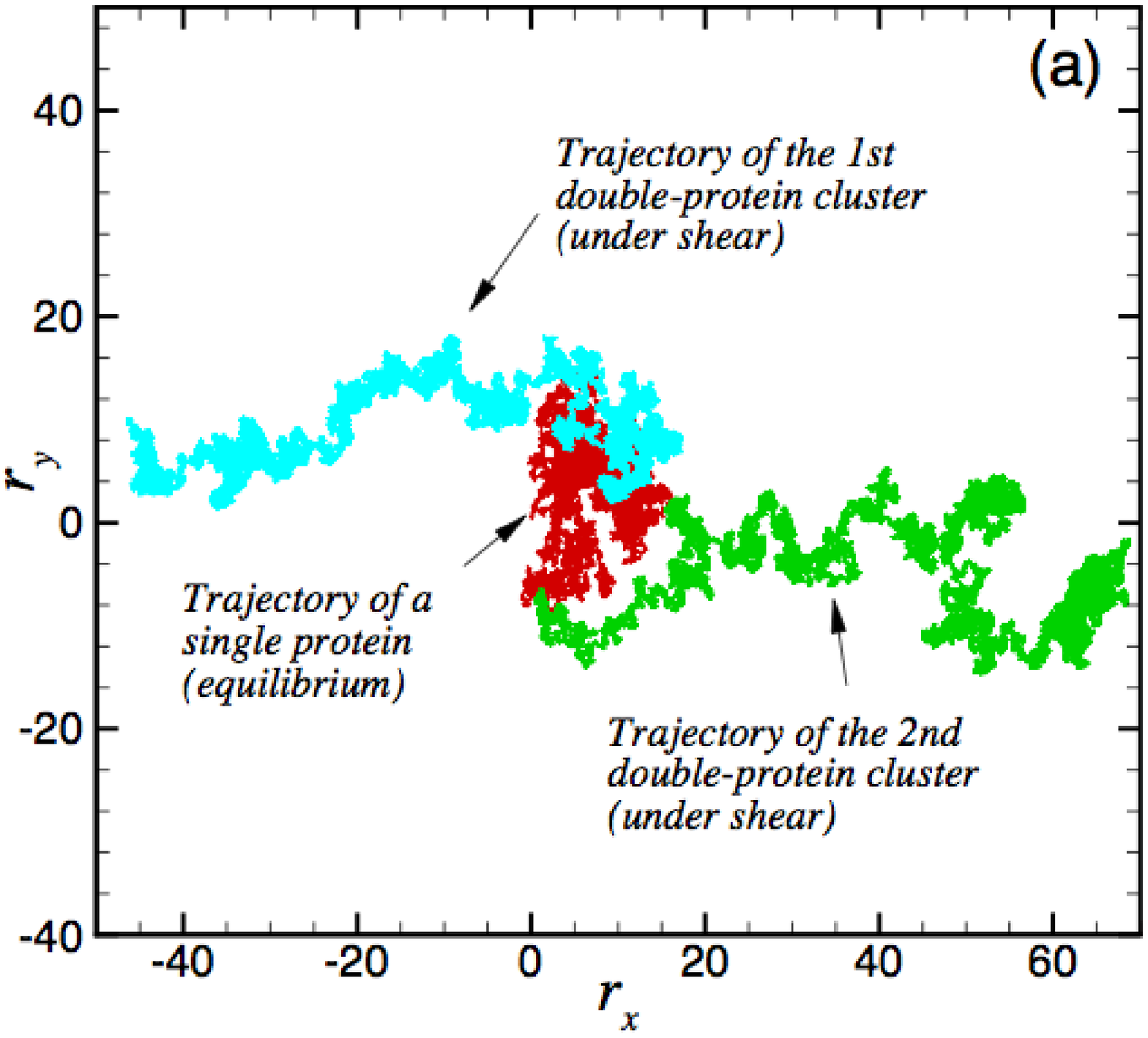} } \hspace{-3mm}
                    \hbox{\includegraphics[width=0.4\textwidth]{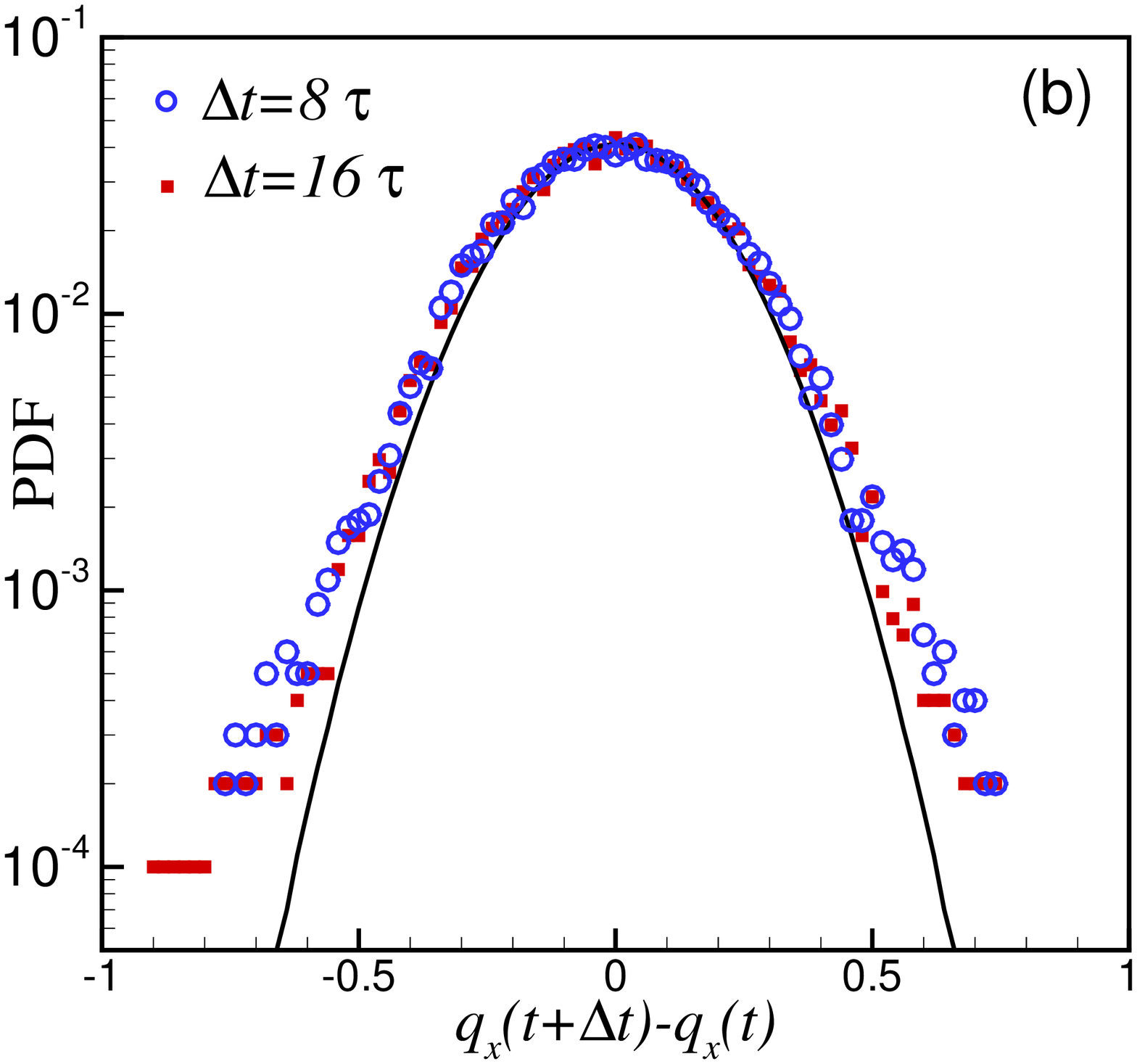} }  }
\centerline{\hbox{\includegraphics[width=0.4\textwidth]{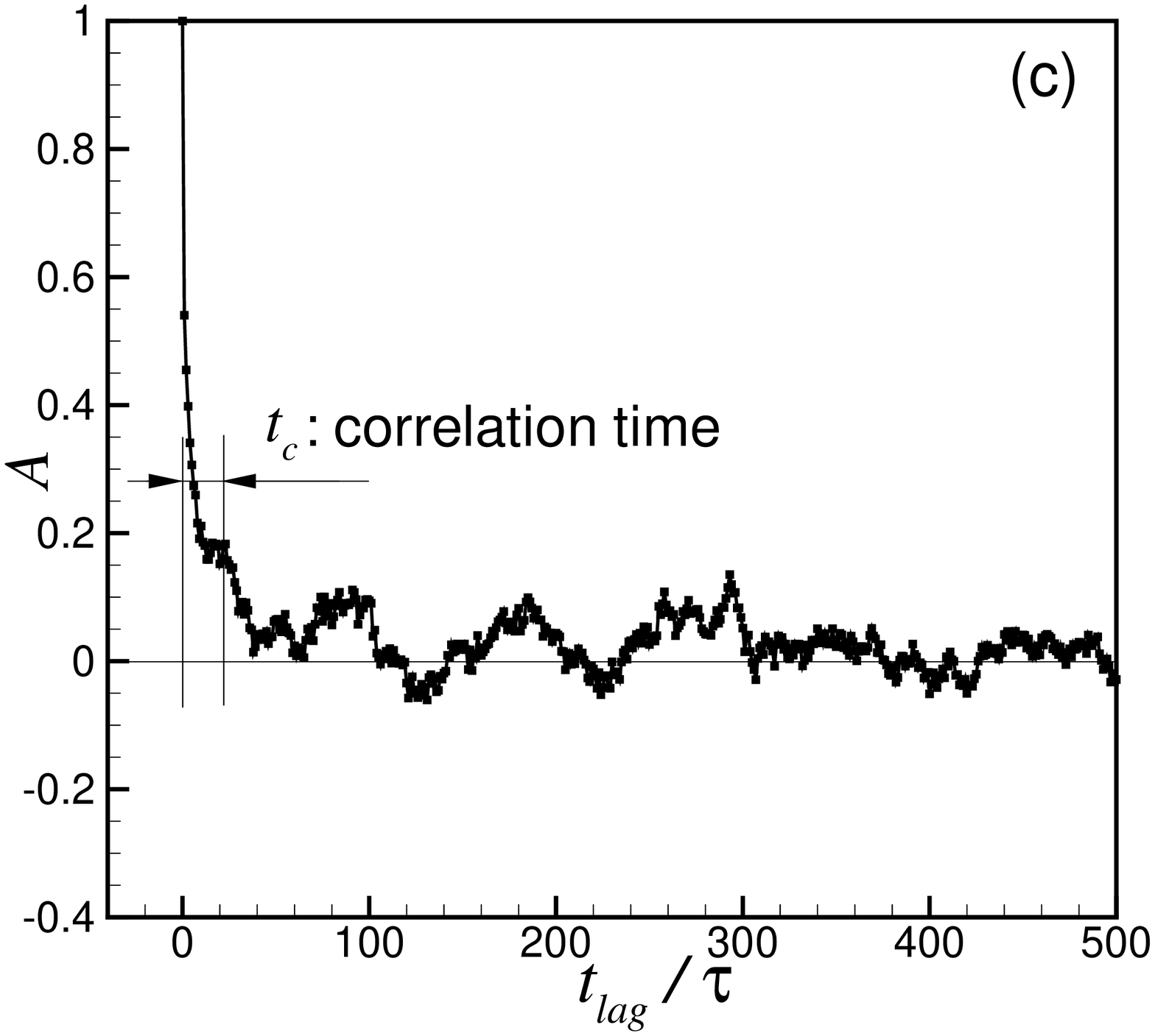} }  \hspace{-3mm}
                     \hbox{\includegraphics[width=0.4\textwidth]{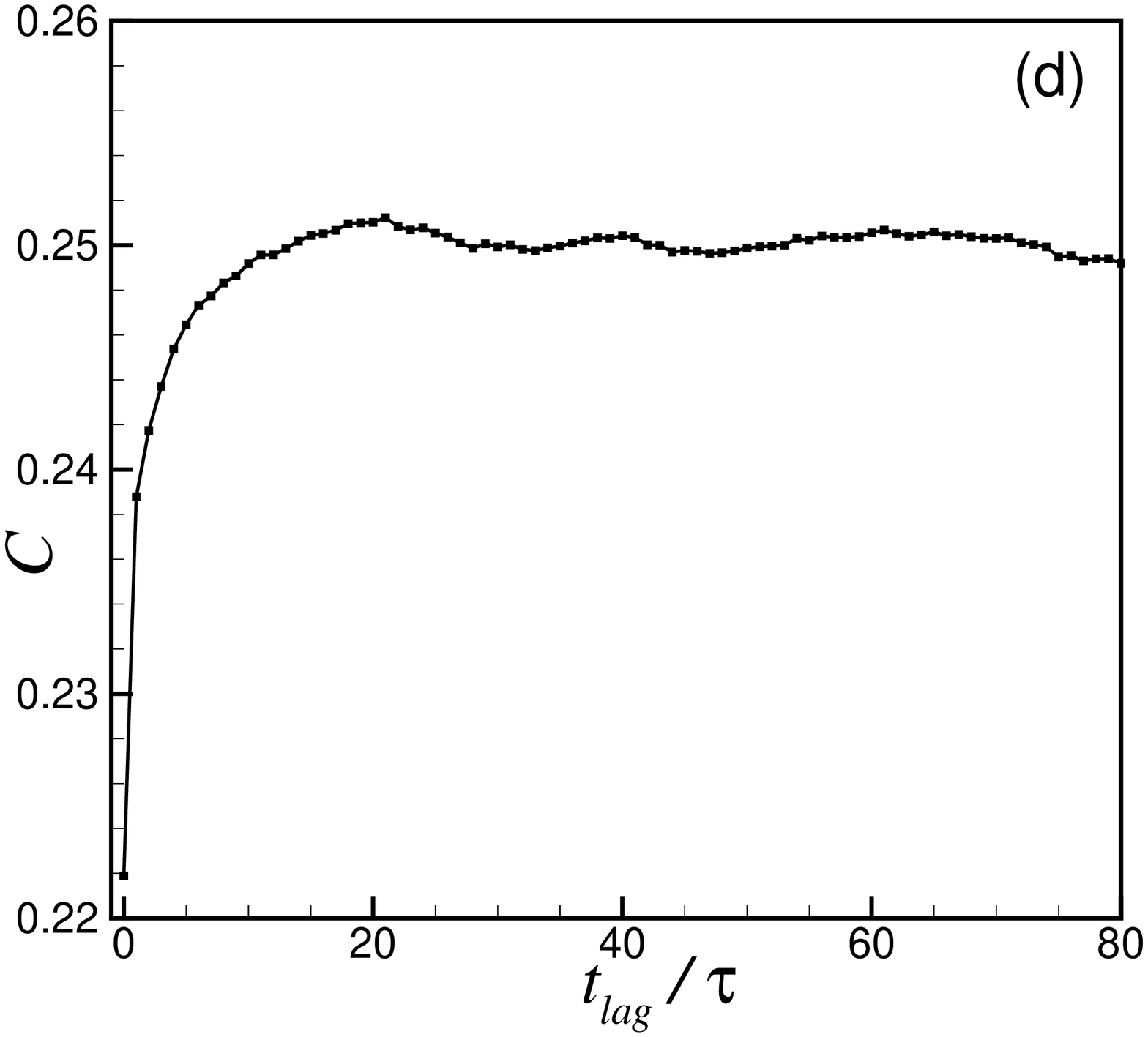} }    }            
\caption{(Color online) (a) Trajectories of a single protein in equilibrium and two double-protein 
clusters under shear flow as pointed by arrows. (b) PDF for the displacements of a sample 
two-protein cluster. Solid line shows the best Gaussian function fitted to the data. 
(c) Autocorrelation function $A(t_{\rm lag})$ for the difference between the head 
particle populations near the two ends of a protein cluster. The correlation time 
$t_c\approx 23\tau$ is defined at the point where $A(t_{\rm lag})$ abruptly drops 
below $10\%$ of its maximum. We have taken 1000 successive samples in the 
time domain, with increments of $1\tau$, to compute $A$. (d) Cross-correlation 
function $C(t_{\rm lag})$ between the distribution of protein--lipid complexes and 
their neighboring voids. The integrals in $C$ have been taken using a grid of 
$29 \times 29$ in the $xy$-plane and 1000 successive points, with steps of $1\tau$,
in the time domain.}
\label{fig4}
\end{figure*}

Fig. \ref{fig4}(a) demonstrates the trajectories of a single protein molecule and two 
double-protein clusters, in equilibrium and sheared systems, respectively. The equilibrium 
trajectory corresponds to regular diffusion because it covers a definite area. In the sheared 
system the trajectories are elongated and aligned with the flow direction, indicating a 
fractional random walk: local isotropic wanderings followed by small-step jumps mainly 
in the flow direction. These successive jumps can be interpreted by inspecting the contour 
plots of $\hat f(\textit{\textbf{r}},\Delta,t)$ over a long duration of time. The hydrophobic 
mismatch between protein clusters (which have asymmetric big cross sections) and the 
membrane disturbs the bilayer thickness and the arrangement of nearby lipids. 
Moreover, proteins are able to transport their neighboring lipids with them and 
behave as dynamic complexes \cite{pra07,nie10}. These two effects collaborate to create 
transient voids whose distribution can be described by $\hat g=1-\hat f$ (light shades 
in Fig. \ref{fig3}). When the bilayer is sheared, protein--lipid groups are pushed into the 
voids created by themselves or other groups/complexes and experience accelerated, 
and therefore, superdiffusive movements. It should be noted that during our simulations, 
the center of mass of the membrane and embedded proteins remains almost at the 
center of the coordinate system.  

None of our samples show long-step straight motions of protein clusters. What we have seen 
are small-step jumps, which are comparable with the mean distance between protein clusters 
and voids [compare Figs. \ref{fig3} and \ref{fig4}(a)]. We have computed the probability 
distribution function (PDF) of protein displacements and plotted it in Fig. \ref{fig4}(b). 
A Gaussian function has been fitted to the data by setting its maximum to the maximum of PDF, 
and its variance is found using the full width at half minimum of PDF. The PDF exhibits 
a deviation from normal distribution and it has tails. We have also applied the Kolmogorov-Smirnov 
test \cite{Press07} to confirm that the PDF is not a normal Gaussian. This is a clear indication 
of anomalous diffusion.

As we noted before, the ends of proteins are pulled in opposite directions by the two 
sheared solvent columns. An important question is why do protein clusters prefer 
to jump into the voids when the membrane is sheared? We have a simple explanation 
for this behavior: Two ends of proteins attract lipids from two different layers of the 
membrane. Thus, the symmetry in the distribution of the upper and lower protein-bound 
lipids is likely to break. Moreover, the shear force is exerted by the solvent on both the 
lipid and protein heads. The mentioned symmetry breaking thus leads to different force 
components at the upper and lower ends of protein--lipid complexes, and they will jump 
into a void in the direction specified by the broken symmetry. To quantify this process,
we take a sample two-protein cluster and define $\rvec_u$ and $\rvec_l$ 
as the center of masses of its protein heads in the upper and lower leaflets, respectively. 
We then compute 
\begin{eqnarray}
\xi(t_i)=\hat f(\rvec_u,\Delta,t_i)-\hat f(\rvec_l,\Delta,t_i),
\end{eqnarray}
which is proportional to the net shear force exerted on the cluster at the time step $t_i$:
the local effective area in contact with a solvent column is determined by the number of 
head particles, and the shear force is calculated by multiplying the effective contact 
area by the shear rate and viscosity. $\xi(t_i)$ will be zero if the concentrations of the 
head particles of lipids are identical around the two heads of the cluster. Defining 
$\bar \xi$ as the average of $\xi(t_i)$, the autocorrelation function 
\begin{eqnarray}
A(t_{\rm lag}) = \frac {\sum_{i} \left [ \xi(t_i+t_{\rm lag})-\bar \xi \, \right ] \cdot 
\left [ \xi(t_i)-\bar \xi \, \right ] }{\sum_i \left [\xi(t_i)-\bar \xi \, \right ]^2},
\end{eqnarray}
plotted in Fig. \ref{fig4}(c) carries interesting information about the shear force experienced 
by the cluster: the correlation time $t_c \approx 23\tau$ shows a sustained accelerated 
motion of the cluster over $t_0 < t \lesssim t_0+t_c$, independent of the initial time $t_0$. 
Moreover, the oscillatory decaying profile of $A(t_{\rm lag})$ shows a random symmetry 
breaking in the sense of deterministic chaos \citep[][\S 5.3.4]{A94}. The existence of a 
correlation time can also be verified by studying the cross-correlation function
\begin{eqnarray}
C(t_{\rm lag})=\int dt \int d\rvec \, \hat f(\rvec,\Delta,t+t_{\rm lag}) \cdot \hat g(\rvec,\Delta,t)
\end{eqnarray}
over one of the leaflets. Fig. 4(d) shows that $C(t_{\rm lag})$ of the upper leaflet steeply rises 
from $t_{\rm lag}=0$ and peaks at $t_{\rm lag} \approx 21\tau$, which is the earliest time span 
that protein--lipid complexes need to occupy their nearest voids. This is quite consistent with 
the acceleration time scale of protein clusters predicted by the autocorrelation function 
$A(t_{\rm lag})$. For $t_{\rm lag} \gtrsim 21\tau$ the cross-correlation function remains 
almost flat because the size of a void is always bigger than the distance that a protein 
cluster travels during $t \sim {\cal O}(t_c)$. \\  
\section{Conclusions}
In this work, we use a toy model of proteins and lipids to simulate the dynamics 
of cell membranes undergoing shear flow. We calculate the MSD profile of lipids 
and proteins in equilibrium and sheared system and compare our result with existing 
works in the literature. This simple model beautifully captures the basic regimes 
in the diffusive behavior of lipid molecules: short initial ballistic regime, transient 
subdiffusive regime and final Normal diffusion. All-atom MD simulations show the 
same diffusive regions in the MSD profile of lipid molecule \cite{fle09}. Moreover, we show 
that shear flow reduces the tension of membrane and consequently decreases the diffusion 
coefficient in the direction of flow. The fact that reducing membrane tension slows down 
lipids movement has been reported as the result of all-atom MD simulations \cite{mud11} 
and experimental measurements \cite{but01}. 

Since there were small number of protein molecules in the system, we did extensive MD 
simulations to obtain the final MSD profiles for proteins. Our C++ code, which has already 
been calibrated \cite{khosh10} for membranes under simple shear flow, is not parallel and allows 
only for small-scale simulations. We have recovered our results by LAMMPS for models 
in equilibrium conditions. However, LAMMPS has not the capability of imposing simple shear 
flow conditions and we could not use it to run our sheared systems over longer time scales.
Despite these limitations, the smooth MSD profiles obtained from our numerous samples clearly 
show the distinction between the regular diffusion of single proteins and anomalous diffusion 
of protein clusters. We explain this anomaly by deliberately examine the distribution of head 
particles of lipids and proteins and introducing void generation mechanism. Our simulations 
cover timescales of order of nanoseconds. 
         
Cell responses to stimuli are fast due to enhanced mobility of protein receptors. 
Consider our findings, superdiffusion of proteins under shear flow can play a 
dominant role in the process of signaling in endothelium cells, RBCs, liposomes 
used for targeted drug delivery and other sheared membranes. We note that since 
the length and shape of proteins and their ability in attracting neighboring lipids 
controls the sizes of voids ---there was no other mechanism in our models for void 
generation--- superdiffusion properties reported in this study are not universal 
and they highly correlate with the properties of embedded proteins. Simulations 
using detailed structures of lipids and proteins are needed to better assess the 
superdiffusive behavior in realistic cell membranes. Experimental exploration 
of our results can be done using single particle tracking \cite{cra08} which can 
give the MSD of proteins directly.\\
\section*{Acknowledgments} 
We thank anonymous referees for their insightful comments that helped us to improve the 
presentation of our results. A.K. was partially supported by the National Elites 
Foundation of Iran. We express our sincere thanks to Prof. Ali Meghdari for 
his helpful discussions and encouragements.

\end{document}